# Radioactive Ion Beam Physics and Nuclear Astrophysics in China

Y.G. Ma[4], X.Z. Cai[4], W.Q. Shen[4]
W.L.Zhan[1], Y.L.Ye[3], W.P.Liu[2], G.M.Jin[1], X.H.Zhou[1]
S.W.Xu[1], L.H.Zuo[2], S.J.Zhu[5], Z.H.Liu[2], J.Meng[3]

[1]*Institute of Modern Physics, CAS, Lanzhou 730000, China*
[2]*China Institute of Atomic Energy, Beijing 102413, China*
[3]*Beijing University, Beijing 100871, China*
[4]*Shanghai Institute of Applied Physics, CAS, Shanghai 201800, China*
[5]*Tsinghua University, Beijing 100871, China*

Abstract

Based on the intermediate energy radioactive Ion Beam Line in Lanzhou (RIBLL) of Heavy Ion Research Facility in Lanzhou (HIRFL) and Low Energy Radioactive Ion Beam Line (GIRAFFE) of Beijing National Tandem Accelerator Lab (HI13), the radioactive ion beam physics and nuclear astrophysics will be researched in detail. The key scientific problems are: the nuclear structure and reaction for nuclear far from β-stability line; the synthesize of new nuclides near drip lines and new super heavy nuclides; the properties of asymmetric nuclear matter with extra large isospin and some nuclear astro- reactions.

## I. Introduction

Recently the development of the radioactive ion beam (RIB) technique has stimulated the research of structure and reaction for exotic nuclei both experimentally and theoretically [1-3]. The studies using RIB demonstrated that a large enhancement of the total reaction cross section ($\sigma_I$) induced by neutron rich nuclei was found which was interpreted as neutron halo[4-7] ( such as $^{11}$Li, $^{11,14}$Be and $^{19}$C etc ) and neutron skin ( such as $^{6}$He and $^{8}$He ) structure[4]. The halo structure of $^{11}$Li seems to be consistent with all the experimental results including the enhancement of $\sigma_I$, the enhancement of two-neutron removal cross section $\sigma_{-2n}$ and the narrow peak in the momentum distribution of fragmentation $^{9}$Li. The discovery of these unusual phenomena evokes further theoretical and experimental research on proton halo and proton skin. New properties of these nuclei like the soft giant resonance, the change of magic number and new decay modes etc stimulate the research using RIB strongly.

RIB research opens the new research area called isospin physics. Isospin dependence of collective flow, particle and intermediate mass fragment (IMF) emission yield ratio and double yield ratio of isotopes etc in intermediate energy heavy ion reaction using different isospin reaction systems is closely related to equation of state(EOS) of nucleus and nucleon-nucleon cross section in nuclear medium which are important for both nuclear physics and astrophysics. RIB will help to synthesize new nuclides far from stability line and even super heavy elements. RIB also stimulates new research area called nuclear astrophysics strongly. The most important base of these researches is the production of RIB with high intensity and good quality. In the past decades the isotope separation on line (ISOL) and projectile fragmentation (PF) methods have proved to be the very efficient ways to produce RIB. Recently a proposal which combines above two ways together suggest to built a new RIB machine which can give four different energy RIB at same time. With the development of RIB technology the radioactive ion beam physics and nuclear astrophysics will obtain new development in near future.

In this paper, the new progresses of Radioactive Ion Beam Physics and Nuclear Astrophysics in China are reports, which focus on the following aspects: the nuclear structure and reaction for nuclear far from β-stability line, the synthesize of new super heavy nuclides; Isospin dependence of High spin state and some systematic related theoretical research.

## II. Halo and Skin Nuclei: Structure and Reaction

Systematic measurements of $\sigma_R$'s for some A<30 proton rich nuclei were performed on RIBLL of HIRFL at intermediate energies[8-10]. The $\sigma_R$ was measured by a direct transmission method using a Si detector telescope, where the radioactive ion beams were produced through the projectile fragmentation induced by 69 MeV/nucleon 36Ar primary beam. As shown in the isospin (N −Z) dependences of $\sigma_R$ for N = 8, 10, 12 isotones in Fig.1 (right), the experimental $\sigma_R$ of $^{23}$Al and $^{27}$P are anomalously larger than their isotones. The difference factor d is used in order to analyze the possibility of nucleon halo or nucleon skin in those nuclei quantitatively. Ozawa et al. [11] had successfully applied the difference factor d to analysis the exotic structure in neutron-rich nuclei. The (N −Z) dependence of d is also given in Fig. 1 (left). The d's of $^{17}$F, $^{23}$Al and $^{27}$P show a large enhancement compared to their neighboring nuclei.

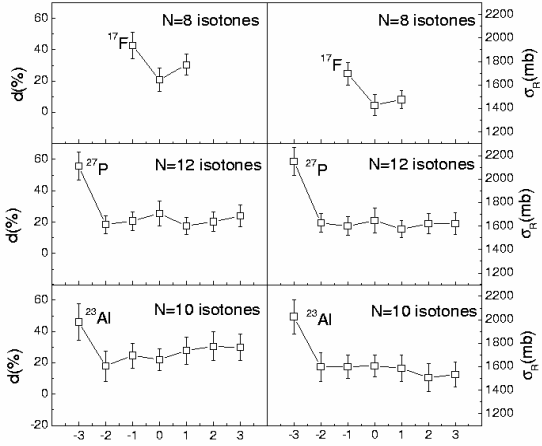

Fig. 1 The (N-Z) dependences for the different factor d and $\sigma_R$ for N = 8, 10, 12 isotones from the present experiment at 30 MeV/u.

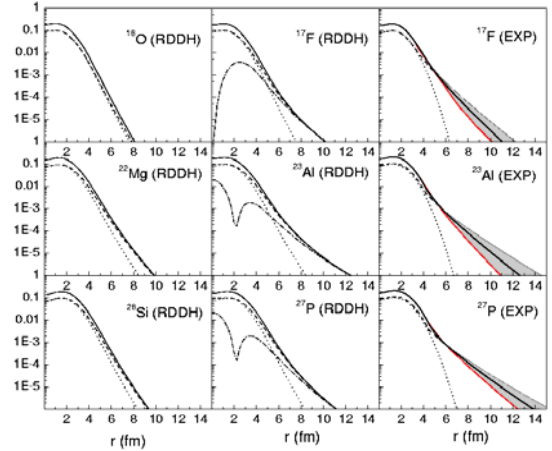

Fig. 2 RDDH calculated Density distributions of proton (dashed), neutron (dotted), matter (solid) and the last proton (dotted dashed) in $^{17}$F, $^{23}$Al, $^{27}$P and that of their corresponding core nuclei. The experimental density distribution deduced using Glauber model for $^{17}$F, $^{23}$Al and $^{27}$P are also plotted as shadow region.

For $^{27}$P, it was predicted [12-13] that there is a proton halo in $^{27}$P and the MSU experiment [14] confirmed it. Together with the data of the separation energy of the last proton ($S_p$ = 0.6 MeV) and of its ground state information (2s1/2), it is strongly suggested that there is a proton halo in $^{27}$P. The calculated result of relativistic density-dependent Hartree (RDDH) approach in Fig. 2 also shows there is a long tail in the proton density distribution. The theoretical RMS radius of the last proton is 4.34 fm and it is evidently larger than the average RMS radius of other nuclei. For $^{23}$Al. the proton separation energy is $S_p$ = 0.125 MeV. Although the spin and parity of the ground state in $^{23}$Al is still unknown, it is known that there is a large prolate deformation in its neighboring nucleus 22Mg ($\beta_2$= 0.56). It is reasonably assumed that the deformation in $^{23}$Al is close to that in $^{22}$Mg. It is very safely concluded that the deformation in $^{23}$Al should be $\beta_2$ = 0.3~0.6. According to the deformed calculations by Bohr and Mottelson [15], it is reliably assumed that the ground state of $^{23}$Al is the $2s_{1/2}$ state. Therefore it is concluded that there is a proton halo in $^{23}$Al. The RDDH calculation on the density distributions of $_{23}$Al (see Fig. 2) also supports a proton halo in $^{23}$Al. For $^{17}$F, the proton separation energy is Sp = 0.6 MeV and its ground state is a spherical state 1d5/2. It may have a proton-skin structure which is also suggested by other experimental group based on various experimental results [16]. Further experiments are needed for proton-rich nuclei in this mass region in order to the elucidation the proton halo structure in $^{23}$Al and $^{27}$P and the proton-skin structure in $^{17}$F.

The origin of the chemical elements in nature constitutes a fascinating problem of nuclear astrophysics. Stellar nucleosynthesis has become one of the most successful theories of element formation[17-18]. In the theory, it assumes that only hydrogen, helium, and the rare light isotopes with A<12 could be produced in the big bang. The nuclei heavier than iron can only be created by successive neutron capture reactions and beta decay. Along this sequence, the $^{11}$B(n, γ)$^{12}$B and $^{12}$C(n, γ)$^{13}$C are very important and therefore experimental studies of these reactions are strongly called for. However, the direct capture cross sections at stellar energies are very difficult to measure with high precision. Recently, an indirect method was proposed to obtain the direct capture cross section at stellar energy from the asymptotic normalization coefficient (ANC) of the overlap wave function in peripheral transfer reaction[19]. The cross sections of the peripheral nucleon transfer at energies above the Coulomb barrier are several orders of magnitude larger than that of direct capture. Hence their method should provide an easy and reliable way to determine the capture cross sections of astrophysical interest.

The experiment was carried out with a collimated deuteron beam from the HI-13 tandem accelerator at China Institute of Atomic Energy, Beijing (CIAE). The differential cross sections of the transfer reactions $^{11}$B(d,p)$^{12}$B and $^{12}$C(d,p)$^{13}$C were measured in the angular range of 5°< $\theta_{Lab}$<140° in 5° steps. The measured angular distribution of the transfer reactions $^{11}$B(d,p)$^{12}$B is shown in Figs. 3 as solid points. Using a proper distorted-wave Born approximation (DWBA) code[20], the nuclear ANC's of the ground and excited states in $^{12}$B and $^{13}$C are derived from the transfer differential cross sections at very forward angles. It shows that the ANC value is insensitive to the parameters used in the binding potential. With these ANC's, we have calculated the rms radii, the probability D1 of the last neutron outside the range of nuclear interaction, and the contribution, D2 of the asymptotic part to the rms radius. The results indicate that the second ($J^{\pi}=2^-$) and third ($J^{\pi}=1^-$) excited states in $^{12}$B and the first ($J^{\pi}=1/2^+$) excited state in 13C are the neutron halo states, whereas the third ($J^{\pi}=5/2^+$) excited state in 13C is a neutron skin state[21]. The ANC method provides a natural way to relate the peripheral transfer reactions to the direct radiative capture reactions. It is of astrophysical interest to deduce the direct neutron radiative capture cross sections of $^{11}$B and $^{12}$C from the measured ANC's. These stable nuclei can also be investigated without radioactive beams. Therefore more and more laboratories can study neutron halos even if they do not have a big machine for the production of radioactive beams. This may be important for future study.

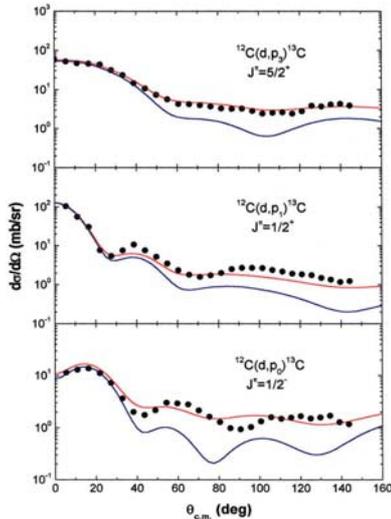

FIG. 3. Angular distributions for the transfer reactions $^{11}$B(d,p)$^{12}$B at $E_d$=511.8 MeV. The dashed curves denote the normalized DWBA calculations. The solid lines represent the results combining the contributions of the direct and the compound processes.

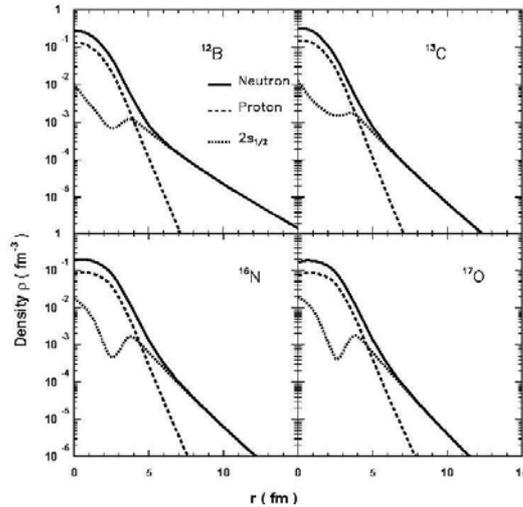

Fig. 4 The matter density distributions for nuclei $^{12}$B, $^{13}$C, $^{16}$N, and $^{17}$O where the last neutron is in the excited state $2s_{1/2}$ with the force set SL3. The solid curve is for the neutron density distribution, the dashed one for the proton density distribution, and the dotted one for the halo neutron, occupying the excited $2s_{1/2}$ state.

Based on the relativistic mean-field (RMF) theory, the newly discovered neutron halos in the excited states of nuclei $^{12}$B, $^{13}$C and $^{209}$Pb [21] are studied. Because these nuclei are spherical with a good inert core plus a valence neutron, the spherical RMF code is used to carry out numerical calculations[12,23]. Here two typical forces sets of effective forces NL-SH and NL3[24-25] are used for and it is expected that other forces will lead to very similar results. The total energy, single particle energy, radii, wave-functions of each nucleon are obtained and the rms radius of the last neutron can be calculated by its wave function. The results show that the theoretical binding energies for the ground state and the excited states are very close to the experimental data. It shows the RMF model can be applied for the lowly excited states of odd-A nuclei with a core plus a valence neutron structure. It is seen clearly that the last neutron is tightly bound for the ground state and weakly bound for the excited state. Therefore there are possibly the neutron skin for the ground state and the neutron halo for the excited state. The RMS radius of the last neutron in the ground state of a nucleus is slightly larger than the matter RMS radius of the nucleus and its core nucleus. But the RMS radius of the last neutron in the excite state is significantly larger than the RMS radius of the nucleus and its core nucleus. This demonstrates the existences of neutron skin in the ground state and neutron halo in the excited state. The self-consistent RMF calculated density distributions for nuclei $^{12}$B, $^{13}$C, $^{16}$N, and $^{17}$O where the last neutron is in the excited state s1/2 are plotted in Fig 4. It is seen that there is a long tail in the density distribution of neutrons due to the weak binding of the last neutron. This clearly indicates that the neutron halo exist in the excited state of these nuclei. We suggest that the $^{16}$N and $^{17}$O can be chosen as the new candidates of neutron halos in the excited state. The present experimental conditions in CIAE are suitable to observe the neutron halo in these two nuclei.

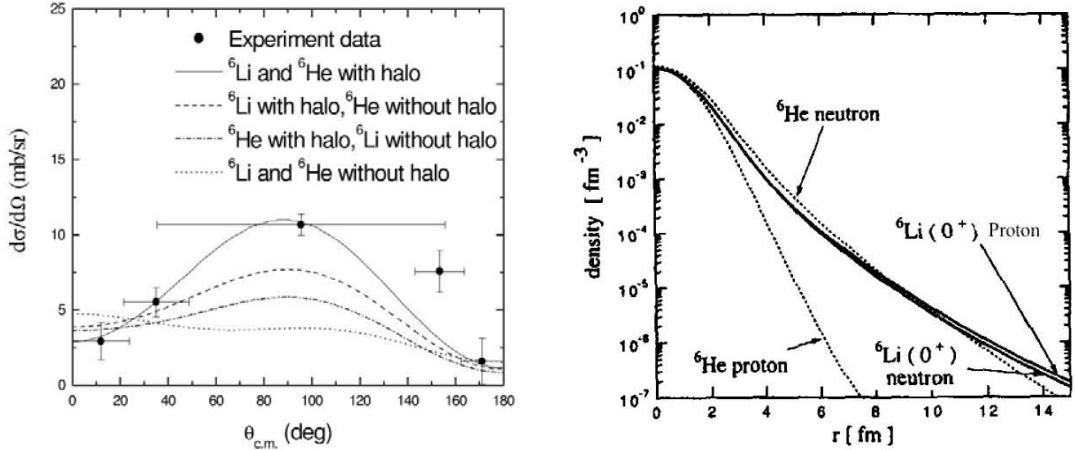

Fig. 5 (a) Angular distributions of $^1$H ($^6$He, $^6$Li)n reactions and the DWBA calculations by different nucleon density distributions. (b) Neutron and proton density distributions for the 3.563 MeV 0+ state of 6Li and the ground state of 6He used in the calculation.

The isobaric analog states (IAS) of the neutron halo nuclei should also have halo-like structure[26]. Arai et al. [27] studied $^6$He and of it is isobaric analog state (3.563 MeV $0^+$ state of $^6$Li) with a fully microscopic three-cluster model and predicted that the latter has a more conspicuous halo-like structure formed by the neutron and the proton surrounding the core. Though the halo effect should appear in the region of large angles, the measurement of full angular distributions for the $^1$H($^6$He, $^6$Li)n reaction in reverse kinematics is necessary. The experiment was carried out using the secondary beam facility[27] of the HI-13 tandem accelerator at CIAE. The angular distributions are shown in Fig. 5(a). The filled circles are the differential cross sections of 3.563MeV $0^+$ state of $^6$Li. The conventional DWBA calculations of the 3.563 MeV $0^+$ state of $^6$Li with different nucleon density distributions are also shown in Fig. 5(a). The experimental data can be well reproduced with the microscopic DWBA analysis if it is assumed that both the ground state of $^6$He and the secondary excited state of $^6$Li have halo structure with nucleon density distributions as shown in Fig. 5(b). It reveals the proton–neutron halo structure of the secondary excited state of $^6$Li which is predicted by Arai et al. [26].

**III. Synthesis of New Nuclides and Super Heavy Nuclides**

Early attempts to discover isotopes of the element Db105 were made by Flerov in 1968 [28]. Up to now, a blank position for 259Db is still left among the known isotopes. Its decay properties are still completely unknown. Identification and study the decay properties of $^{259}$Db nuclide will close this gap[29]. The reaction $^{241}$Am ($^{22}$Ne, 4n) was used to produce this new isotope $^{259}$Db, which is performed at the SFC (Sector Focus Cyclotron) of HIRFL. Its identification was performed by recoil-milking the 21s $^{255}$Lr daughter. At the same time, its neighboring nuclide $^{258}$Db has also been observed. The experimental set-up is shown schematically in Fig. 6(a).

The identification of this nuclide has been performed by measuring the alpha-particle emission of the mother and daughter nuclides. A complex group of peaks with the energies of 9.08, 9.17 and 9.30 MeV could be assigned to $^{258}$Db based on the whole complex group decays with a measured half-life of 4.3s. This nuclide arose from the reaction $^{241}$Am ($^{22}$Ne, 4n). According to Alice Code calculations, we expect that at E = 118 MeV the 5n cross-section is about a factor of two higher than the 4n cross-section. The half-lives of 4.3 s and 13.4 s deduced from the decay curves of $^{259}$Db and $^{254}$Lr agree quite well with the known values for these isotopes. The production cross-sections estimated from the yields of $^{259}$Db and $^{258}$Db α-decays are 1.6±1.2nb and 3.6±1.8nb, respectively.

The α-decay energies $Q_\alpha$ of the heaviest elements as a function of neutron number N are plotted in fig. 6(b)[29]. We compared the α-decay energies of new isotope $^{259}$Db to the values of known isotopes in a "$Q_\alpha$-systematics" for isotopes with Z ≥ 98. One can see that the deformed gap of N = 152 in the single-particle levels is still observed from elements 98 up to 105. It also shows that the $Q_\alpha$ value for new isotope $^{259}$Db, as compared with the trend for the other Z = 105 isotopes, fits quite well into the general trend. 259Db with neutron number N = 154 has a higher α-decay energy than the other Z = 105 isotopes. Our $Q_\alpha$ value 9.62 MeV for the new isotope $^{259}$Db derived from the experiment is in good agreement with the values of 9.60 and 9.61 MeV theoretically predicted, respectively.

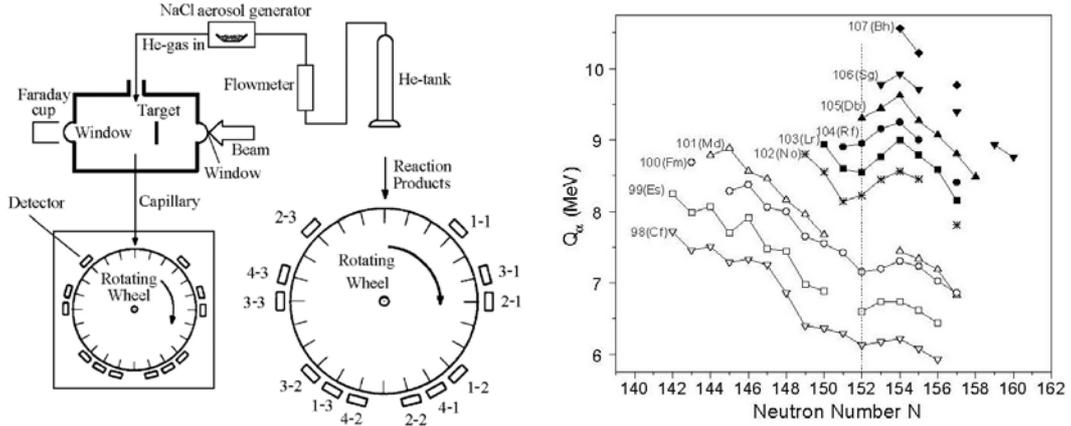

Fig. 6 (a) Schematic diagram of the He-jet target chamber, capillary transport assembly, the rotating wheel collection and detection system. (b) The systematic of alpha-decay energy $Q_\alpha$ vs. neutron number N for isotopes with Z ⩾ 98. The $Q_\alpha$ value for $^{259}$Db was derived from the present work.

**IV. Isospin Dependence of High Spin State**

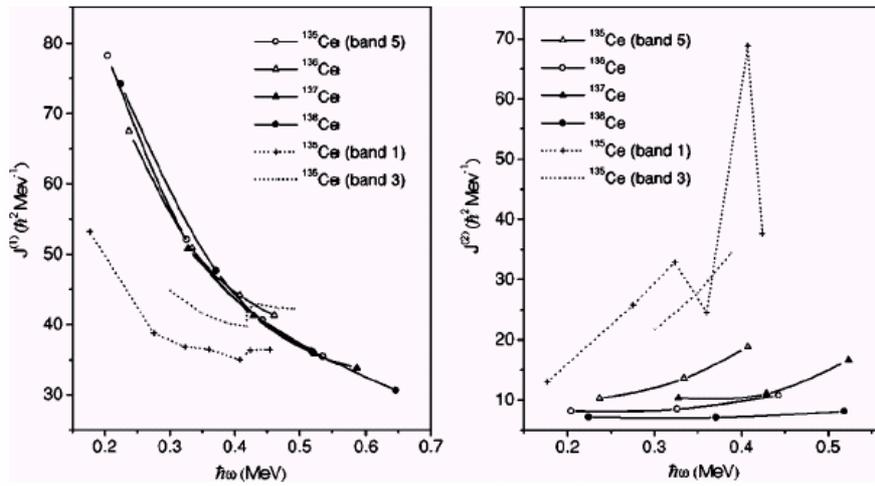

FIG. 7 Comparison of the moments of inertia $J^{(1)}$ and $J^{(2)}$ of the collective band in $^{137}$Ce with those of bands 1, 3, and 5 in $^{135}$Ce and negative parity collective bands in $^{136,138}$Ce.

The $^{137}$Ce nucleus exhibits transitional character as the neutron number is just below the neutron shell closure at N=82. For transitional nuclei in the N=135 region, the ground state shape changes rapidly with N. Study of the high spin states of $^{137}$Ce can provide valuable information for the nuclear structure, the systematical shape change, the quasi-particle configurations of certain single energy levels, and the deformation driving effects of the single particle orbital. The results of in-beam spectroscopic studies of $^{137}$Ce[30-32] indicate that the yrast levels of $^{137}$Ce at low spins show a $h_{11/2}^{22}$ decoupled band structure and the prolate-oblate shape transition may occur between N=77 and 79 in Ce isotopes. However, no high K collective rotational band based on the multi-quasi-particle configuration was observed in $^{137}$Ce. In order to obtain further information on the behavior of the Ce isotope, high spin states in the N=79 $^{137}$Ce nucleus have been studied with the reaction $^{124}$Sn($^{18}$O,5n)$^{137}$Ce at beam energy 78 MeV[33].

The level scheme has been expanded with spin up to 43/2 $\hbar$. At low spins, the yrast collective structure built on the $\nu h_{11/2} \times 2^+$ multiplet shows a transitional shape with $\gamma=30.8°$ according to calculations of the triaxial rotor-plus-particle model. The configurations for several high spin states have been discussed from a systematical comparison with neighboring nuclei. A collective band built on I=33/2 level has been observed for the first time. The most interesting finding of the present study is the collective rotational band built on the 5379.1 keV level. Based on a systematical comparison with the neighboring nuclei $^{135,136,138}$Ce, we believe that this collective band should belong to a collective rotational oblate band built on a multi-quasi-particle configuration with $\gamma=-60°$. The collective band observed in $^{137}$Ce exhibit some distinct properties of the oblate bands in this region. Plots of the moments of inertia $J^{(1)}$ and $J^{(2)}$ of the collective band in $^{137}$Ce along with bands 1, 3, and 5 in $^{135}$Ce and the negative parity bands in $^{136,138}$Ce against the rotational frequency $\hbar\omega$ are shown in Fig. 7. Among them, bands 1 and 3 in $^{135}$Ce have a prolate shape ($\gamma=0°$) and the other three bands in $^{135,136,138}$Ce have a oblate shape ($\gamma=-60°$). The J(1) and J(2) of the band in $^{137}$Ce show a similar behavior to that in the oblate bands of the neighboring isotopes, but are different from those in the prolate bands. Furthermore, the possible configuration for the oblate band in $^{137}$Ce can be deduced which is same as that in band 5 in $^{135}$Ce and consistent with that of the negative parity band in $^{136,138}$Ce.

## V. Theoretical Research and Calculation

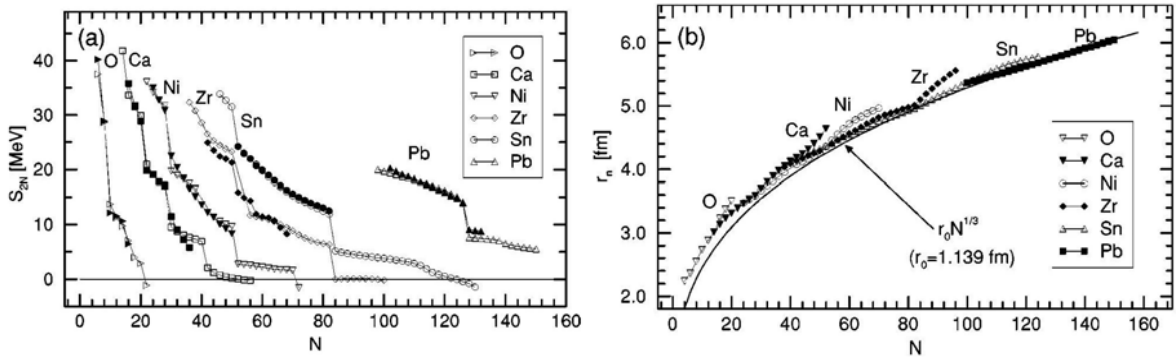

FIG. 8 (a) The two-neutron separation energies $S_{2n}$ in even Ca, Ni, Zr, Sn, and Pb isotopes are plotted against the neutron number N. Open symbols represent the values calculated from the RCHB theory with the NLSH parameter set while solid ones represent the data available. (b) The root mean square neutron radius $r_n$ for even Ca, Ni, Zr, Sn and Pb isotopes from RCHB calculation as a function of the neutron number N. The curve for the r0N1/3 rule with r0=1.139 fm has been included to guide the eye.

Based on the relativistic continuum Hartree-Bogoliubov (RCHB) theory[34], the formation of a neutron halo can be understood as the scattering of Cooper pairs into the continuum containing low-lying resonances of small angular momentum. Along this line a new phenomenon -giant neutron halos- has been predicted in the Zr nuclei close to the neutron drip line[35]. They are formed by from two to six neutrons scattered as Cooper pairs mainly to the levels $3p_{3/2}$, $2f_{7/2}$, $3p_{1/2}$, and $2f_{5/2}$. These procedures are applied in the O, Ca, Ni, Zr, Sn, and Pb isotopes, where the RCHB equations is solved in a box with the size $R$=20 fm and a step size of 0.1 fm. The parameter set NL-SH is used, which aims at describing both the stable and exotic nuclei. The use of other parameters such as *TM*1 does not provide very different results.

The two-neutron separation energy $S_{2n}$ is quite a sensitive quantity to test a microscopic theory. Both the theoretical and the available experimental $S_{2n}$ for O, Ca, Ni, Zr, Sn, and Pb isotopes are presented in Fig. 8(a). The good agreement between experiment and calculation is clearly shown. Along the $S_{2n}$ versus N curve for Ca isotopes, three strong kinks appear at the magic or sub-magic numbers N=20, 28, and 40, respectively. However, there seems no kink at the other magic number N=50, which suggests the disappearance of the neutron magic number N=50. Therefore the nucleus $^{70}$Ca is no more a double-magic nucleus. This disappearance of the N=50 magic number at the neutron drip line is due to the halo property of the neutron density with the spherical shape being kept. All these suggest that the single-particle level energies and their orders will vary violently in the near drip-line region. Then result in the change of the traditional magic numbers. Another remarkable appearance in Fig. 1 is that the $S_{2n}$ values for exotic Ca isotopes are extremely close to zero in several isotopes. If one regards $^{60}$Ca as a core, then the valence neutrons are filled in the weakly bound levels and continuum above the N=40 sub-shell for these nuclei, especially for $^{66\sim72}$Ca. This is very similar to $S_{2n}$ in Zr isotopes with N>82[34]. It should be noted that the behavior of $S^{2n}$ indicates the appearance of a ''giant halo'' in Ca chains [35], just as that in Zr chains.

To illustrate the probable ''giant halo,'' the calculated neutron radii $r_n$ from the RCHB calculation for even-even nuclei in O, Ca, Ni, Zr, Sn, and Pb isotopes are plotted in Fig. 8(b). It is very interesting to see that $r_n$ follow the $N_{1/3}$ systematic well for stable nuclei although their proton number is quite different. Near the drip line, abnormal behaviors appear at N=40 in Ca isotopes and at N=82 in Zr isotopes. The increase of $r_n$ in exotic Ni and Sn nuclei is not as fast as that in Ca and Zr. The nuclei at the abnormal $r_n$ increase correspond to those for $S_{2n}$. They give further support for the formation of a giant halo. Compared with Zr isotopes, the giant halo in Ca isotopes will be easy to access experimentally. Of course there are also some abnormal changes for the Ni and Sn isotopes near the neutron drip line, but they are not as obvious as the Ca and Zr cases.

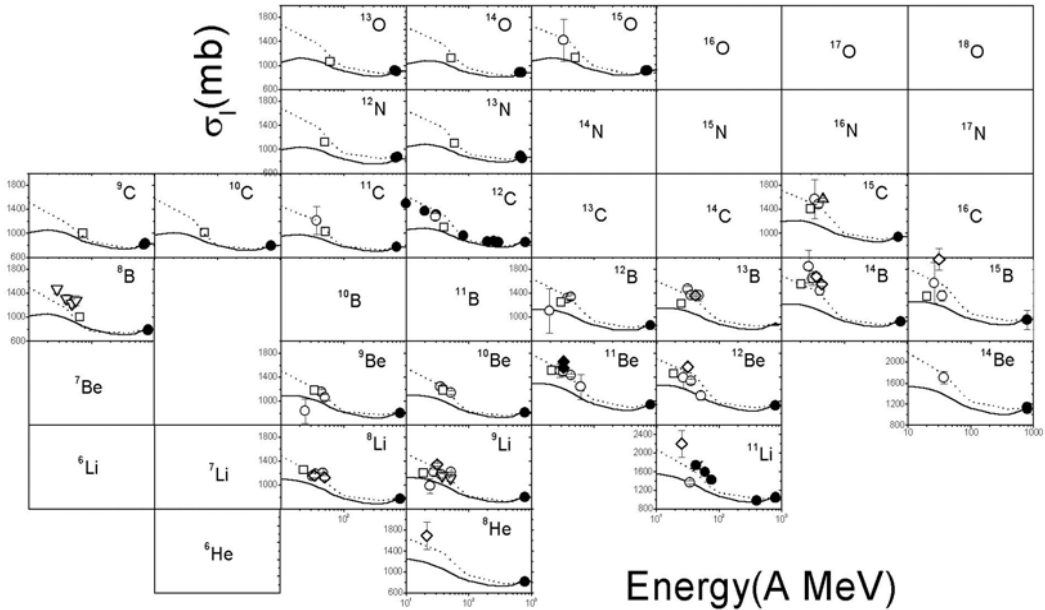

Fig. 9. Incident energy dependence of $\sigma_R$ shown in the of p-shell nuclear chart. The experimental data are indicated by various points[36]. Solid curves indicated the calculations by using the Glauber-model. Dashed curves indicated the BUU calculations by using soft EOS and 0.8 $\sigma_{Cug}$.

A useful tool to study $\sigma_R$ is the microscopic Glauber multiple-scattering theory. Comparisons of the $\sigma_R$ at relativistic energy with that at intermediate energies have been done by Ozawa et al.[11]. It indicates that the Glauber-model calculations always underestimate $\sigma_R$ at intermediate energies. For more detail, the differences between the experimental data and the calculated values at intermediate energies vary from nucleus to nucleus in the same isotope chain, particularly from nuclei near β-stability line to halo or skin nuclei. It has been pointed out that the optical limit may not be a good approximation if one applies the model to halo nuclei at intermediate energy. In order to study this deviation between the experimental data and the calculated one, the Boltzmann-Uehling-Uhlenbeck (BUU) equation has been used to $\sigma_R$ [36]. The $\sigma_R$'s are calculated by using soft EOS and $0.8\sigma_{Cug}$. For simplicity, a square-type distribution for the density is used and the width of the square distribution was obtained by fitting the $\sigma_R$'s at relativistic energies. As shown in Fig. 9, it is very clear that the BUU model reproduces the $\sigma_R$ at intermediate energy much better than Glauber-model.

The systematical underestimation of $\sigma_R$ by Glauber model was removed out now by the above BUU calculation framework. The isospin dependence of $\sigma_R$ has been studied by deducing the difference factor d. From the isospin dependence of d for Li, Be, B and C isotopes, it can be seen that the d of BUU-model calculations is less than that of

Glauber calculations systematically. For Glauber calculations, the d is about 20% for the nuclei near β-stability line and enhances up to 30~40% for some nuclei that have anomalous structure such as skin and halo. For BUU calculations, the d is less than 10% and the d of halo or skin nuclei is still larger than that of their neighbors which is same with the Glauber calculations. Thus it is suggested that the difference factor $d$ is sensitive to the nuclei structure such as halo or skin.

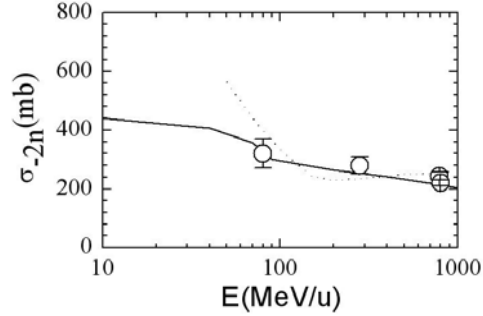

Fig. 10 The energy dependence of $\sigma_{-2n}$ of $^{11}Li+^{12}C$. The open circles are experimental data. The solid curve is the BUU calculation. The dotted curve is the Glauber calculation.

Furthermore the above BUU model is used to study $\sigma_R$、$\sigma_{-n}$ and $\sigma_{-2n}$ simultaneously [37]. The density distributions which come from RMF model with NLZ interaction have been introduced to replace the normal used square-type distributions. Then the BUU calculations, which with no adjustable parameter, can reproduce the experimental $\sigma_R$、$\sigma_{-n}$ and $\sigma_{-2n}$ very well for various reaction systems. Here $\sigma_{-n}$ and $\sigma_{-2n}$ have been calculated as the difference between $\sigma_R$ of halo nucleus and core nucleus, by assuming $\sigma_{corr} \approx 0$. Fig.10 shows the energy dependence of $\sigma_{-2n}$ of $^{11}Li+^{12}C$ reaction system. It can be seen that the BUU calculated $\sigma_{-2n}$ decreases with the increase of energy at both low and high energy range gently. It suggests that the assumption works well at high energy. At intermediate energy, the calculated $\sigma_{-2n}$ of Glauber calculation increases fast with the decrease of energy, where the BUU calculation gives slight increasing trend. Whatever the experimental data of $\sigma_{-2n}$ at this energy range is rather rare. More experimental measurements are necessary to estimate the roles of different mechanisms simultaneously and test the validity of the assumption at intermediate energy.

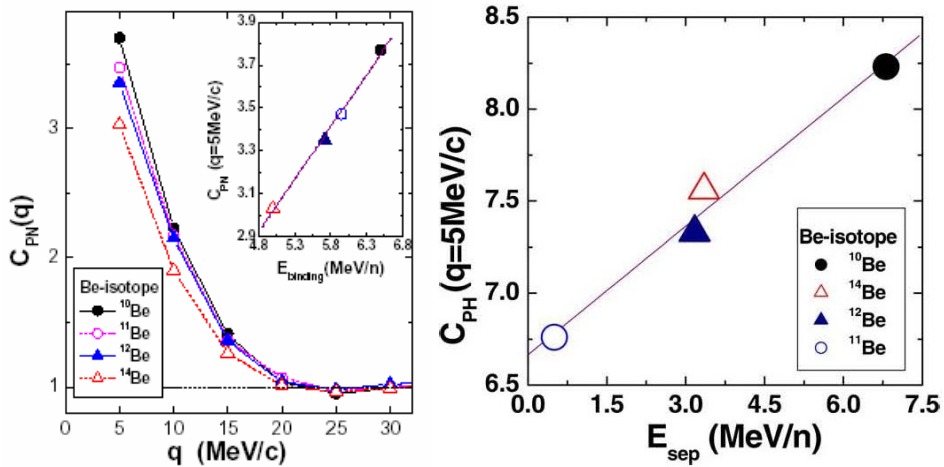

Fig. 11 (a) The proton–neutron correlation function for different Be isotopes. The insert shows the relationship between the strength of proton–neutron correlation function CPN at 5 MeV/c and the binding energy per nucleon of the projectile Eb. (b) The relationship between the proton–halo-neutron correlation function CPH at 5 MeV/c and the single-neutron separation energy of the nuclei. The solid line is just a linear fit.

Hanbury Brown–Twiss (HBT) technique was presented for the astrophysical measurements a few decades ago, and it reveals information about the angular diameters of distant stars. More recently, it has been widely used in others fields since the emission time and source size in the nuclear reaction can be extracted by the nuclear HBT technique. It is very interesting to investigate the exotic nuclei via HBT technique further. The intensity HBT technique has been applied to investigate its sensitivity to the binding energy and separation energy of neutron-rich nuclei from the break-up of nuclei by convoluting the phase-space distribution generated with the IDQMD model. Firstly we gave a well-fitted halo-neutron–halo-neutron correlation function from the break-up of $^{14}$Be on C target. Based upon this achievement of the good fit, we explore the dependence of the proton–neutron correlation function ($C_{PN}$) at small relative momentum with the binding energy ($E_{bin}$) for Be isotopes. It was found that the correlation strength of $C_{PN}$ at small relative momentum rises with $E_{bin}$. This changeable tendency of $C_{PN}$ with $E_{bin}$ is reported for the first time and it might be a potential good way to study the structure of the nuclei. Moreover, the proton–halo neutron correlation function ($C_{PH}$) is also constructed from the break-up reactions. There exists the similar relationship between the $C_{PH}$ at small relative momentum and the separation energy ($E_{sep}$) of Be isotopes as the relationship of CPN to $E_{bin}$. From theoretical point of view, $C_{PH}$ at small relative momentum is sensitive to $E_{sep}$ and this can be attributed to the spatial extension level of the neutron which is most far away the center of the nucleus.

The behavior of the correlation functions between the proton and the neutron as a function of neutron number and binding energy is shown in Fig. 11(a). Generally the strength of HBT at very small relative momentum shows a clear dependence on the neutron number. The tendency of the $C_{PN}$ rises with the increasing Eb. Among the projectiles we studied, the number of the protons is 4 and that of the neutron are gradually increasing, and this will reflect the stability of the nuclei. With the increasing of the mass number, the mean relationship between the nucleons will change weaker. Since the strength $C_{PN}$ of correlation function symbolizes the mean relationship between the emitted proton and neutron and the binding energy per nucleon associated with the tightness between the nucleon, the tendency shown in the insert of Fig. 11(a) reflects that the $C_{PN}$ can reveal the compactness of the nuclei. On the other hand, we also study the correlation functions between the proton and the most outside neutron. Here the most outside neutron is defined as the one which is the most far away from the spatial center at the FSI and it is also called as a halo neutron for simplification, even though it is not strict in physics sense. To obtain reasonable HBT results for halo neutron and proton, only those events in which the halo neutron and proton are emitted in the same event are accepted to investigate such a correlation function. The similar correlation function was obtained as $C_{PN}$ and the relationship between the strength of proton– halo-neutron correlation function $C_{PH}$ at 5MeV/c and the single-neutron separation energy of the projectile $E_{sep}$ was extracted. The symbols of Fig. 11(b) show the calculated result. It looks that, with the increasing of the $E_{sep}$, $C_{PH}$ at 5 MeV/c rises gradually. Since the strength $C_{PH}$ of correlation function between neutron and proton increases with the decreasing of the source size, the above $E_{sep}$ dependence of $C_{PH}$ reflects that the emission source size of the neutron and proton shrinks with the separation energy of the single neutron, which is consistent with the extent of binding of single neutron via $E_{sep}$.

**VI. Conclusion**

Based on RIBLL of HIRFL and Low Energy GIRAFFE of HI13, the radioactive ion beam physics and nuclear astrophysics will be researched in detail. The key scientific problems which will be researched as high priority are: the nuclear structure and reaction for nuclear far from β-stability line; the synthesize of new nuclides near drip lines and new super heavy nuclides; the properties of asymmetric nuclear matter with extra large isospin and some nuclear astro- reactions. The above research was chosen as one program of Major State Basic Research Development Program in year 2000. This program is performed from 2001 to 2005 and is supported by Ministry of Science and Technology of China. This project is divided as 7 major important tasks: halo nucleus; synthesize of new nuclides near drip line and super heavy nuclides; the structure and reaction with neutron rich nucleus; the structure and reaction with proton rich nucleus; the isospin dependence of high spin states; some important nuclear astro-reactions; some systematic theoretical research. It will be the base of scientific goals of National Big  Science Projects: "Cooler Storage Ring (CSR)" in Lanzhou which will be fulfilled at 2005 on the schedule and "Beijing Radioactive Nuclei Beam Facility (BRNBF)".

This project will be supported by Major State Basic Development Program Under Contract Number G2000774.